\documentclass[sn-nature,iicol]{sn-jnl}% Style for submissions to Nature Portfolio journals
\usepackage{graphicx}%
\usepackage{multirow}%
\usepackage{amsmath,amssymb,amsfonts}%
\usepackage{amsthm}%
\usepackage{mathrsfs}%
\usepackage[title]{appendix}%
\usepackage{xcolor}%
\usepackage{textcomp}%
\usepackage{manyfoot}%
\usepackage{booktabs}%
\usepackage{algorithm}%
\usepackage{algorithmicx}%
\usepackage{algpseudocode}%
\usepackage{listings}%
\usepackage{upgreek}
\theoremstyle{thmstyleone}%
\theoremstyle{thmstyletwo}%

\theoremstyle{thmstylethree}%

\newcommand{\mb}[1]{\mathbf{#1}}

\newcommand{\um}{$\upmu$m~}

\begin{document}

\title[Tunable mechanical properties and air-based lubrication in an acoustically levitated granular material]{Tunable mechanical properties and air-based lubrication in an acoustically levitated granular material}

%%=============================================================%%
%% Prefix	-> \pfx{Dr}
%% GivenName	-> \fnm{Joergen W.}
%% Particle	-> \spfx{van der} -> surname prefix
%% FamilyName	-> \sur{Ploeg}
%% Suffix	-> \sfx{IV}
%% NatureName	-> \tanm{Poet Laureate} -> Title after name
%% Degrees	-> \dgr{MSc, PhD}
%% \author*[1,2]{\pfx{Dr} \fnm{Joergen W.} \spfx{van der} \sur{Ploeg} \sfx{IV} \tanm{Poet Laureate} 
%%                 \dgr{MSc, PhD}}\email{iauthor@gmail.com}
%%=============================================================%%

\author*[1,2]{\fnm{Nina M.} \sur{Brown}}\email{nmbrown@uchicago.edu}

\author[2,3]{\fnm{Bryan} \sur{VanSaders}}\email{bv336@drexel.edu}

\author[4]{\fnm{Jason M.} \sur{Kronenfeld}}\email{jkronen@stanford.edu}

\author[5,6]{\fnm{Joseph M.} \sur{DeSimone}}\email{jmdesimone@stanford.edu}

\author[1,2]{\fnm{Heinrich M.} \sur{Jaeger}} \email{jaeger@uchicago.edu}

\affil[1]{\orgdiv{James Franck Institute}, \orgname{University of Chicago}, \orgaddress{\street{5801 S.\ Ellis Ave}, \city{Chicago}, \postcode{60637}, \state{IL}}, \country{USA}}

\affil[2]{\orgdiv{Department of Physics}, \orgname{University of Chicago}, \orgaddress{\street{5801 S.\ Ellis Ave}, \city{Chicago}, \postcode{60637}, \state{IL}}, \country{USA}}

\affil[3]{\orgdiv{Department of Physics}, \orgname{Drexel University}, \orgaddress{\street{3141 Chestnut St}, \city{Philadelphia}, \postcode{19104}, \state{PA}}, \country{USA}}

\affil[4]{\orgdiv{Department of Chemistry}, \orgname{Stanford University}, \orgaddress{\street{450 Jane Stanford Way}, \city{Stanford}, \postcode{94305}, \state{CA}}, \country{USA}}

\affil[5]{\orgdiv{Department of Radiology}, \orgname{Stanford University}, \orgaddress{\street{450 Jane Stanford Way}, \city{Stanford}, \postcode{94305}, \state{CA}}, \country{USA}}

\affil[6]{\orgdiv{Department of Chemical Engineering}, \orgname{Stanford University}, \orgaddress{\street{450 Jane Stanford Way}, \city{Stanford}, \postcode{94305}, \state{CA}}, \country{USA}}

%%==================================%%
%% sample for unstructured abstract %%
%%==================================%%

\abstract{
Cohesive granular materials are found in many natural and industrial environments, but experimental platforms for exploring the innate mechanical properties of these materials are often limited by the difficulty of adjusting cohesion strength.
Granular particles levitated in an acoustic cavity form a model system to address this. Such particles self-assemble into free-floating, quasi-two-dimensional raft structures which are held together by acoustic scattering forces; the strength of this attraction can be changed simply by modifying the sound field.
We investigate the mechanical properties of acoustically bound granular rafts using substrate-free micro-scale shear tests.
We first demonstrate deformation of rafts of spheres and the dependence of this deformation on acoustic pressure.
We then apply these methods to rafts composed of anisotropic sand grains and smaller spheres, in which the smaller spheres have a thin layer of air separating them from other grain surfaces.
These spheres act as soft, effectively frictionless particles that populate the interstices between the larger grains, which enables us to investigate the effect of lubricating the mixture in the presence of large-grain cohesion.
}


\maketitle

\section{Introduction}\label{sec:intro}

Typical granular materials interact primarily via friction, 
while cohesive granular materials have additional interparticle forces that change both local and bulk behavior.
For example, in geophysical mixtures such as wet soil, particles are often connected by forces due to capillary bridges \cite{Ralaiarisoa_Dupont_Moctar_Naaim-Bouvet_Oger_Valance_2022},
fine powders in industrial settings can experience cohesion from van der Waals or electrostatic forces \cite{Castellanos_2005,Kolehmainen_Ozel_Gu_Shinbrot_Sundaresan_2018}, and
 regolith is bound partly by gravity to asteroids in space \cite{Sánchez_Scheeres_2020}.

In investigating cohesive granular materials the strength of cohesive interactions is mainly altered in one of two ways: changing how samples are prepared or changing the magnitude and direction of fields applied to particles with specific materials properties.
Glass beads coated with polymer can be prepared with different interaction strengths by varying the coating thickness, particle size, or stiffness of the coating material \cite{Hemmerle_Schröter_Goehring_2016,Gans_Pouliquen_Nicolas_2020}. 
In this case, tuning cohesion requires the preparation of different batches of particles, and
the interaction strength cannot be changed \textit{in situ} or for the same sample.
An alternative approach makes use of magnetic forces to introduce cohesion \cite{Peters_Lemaire_2004, Lehman_Christman_Jacobs_Johnson_Palchoudhuri_Tieman_Vajpeyi_Wainwright_Walker_Wilson_et_al._2022}.
Here, the cohesion between particles can be altered continuously and \textit{in situ} by changing an externally applied magnetic field.
For example, applying a magnetic field to a sample of ferromagnetic grains can significantly affect the packing structures that form in the bulk and the flow rate \cite{Lumay_Vandewalle_2007, Lumay_Vandewalle_2008}.
However, this approach requires particles made from magnetizable material.

Acoustic levitation offers another alternative. 
Sound can couple to solids of practically any material, size, and shape. 
Acoustic interactions can be attractive as well as repulsive, can easily be tuned by adjusting the strength of the sound field, and, as
 long as the particles are solid, these interactions effectively do not depend on particle composition.
Sufficiently intense sound can lift particles of up to  few millimeters in size against gravity, and sound-induced attractive forces among levitating particles can then assemble them into freely floating granular rafts. 
Such rafts are comprised of a monolayer of particles and thus effectively form a two-dimensional granular model system, which allows for direct observation and tracking of individual particles in a containerless environment, i.e., without being influenced by contact with walls or a substrate.
In addition, we recently developed a method to apply stress or strain to levitating rafts, making it possible to perform a variety of load tests similar to standard materials testing \cite{Brown_VanSaders_Kronenfeld_DeSimone_Jaeger_2024}.  

For acoustic levitation in air, ultrasound in the range 30-100 kHz is typically used. 
Under these conditions, solid particles larger than about 70 $\upmu$m experience sound-induced forces that are primarily attractive, while smaller particles also experience significant repulsive forces \cite{Lim_VanSaders_Jaeger_2024}. 
The smaller `fines' can therefore interact without making contact, opening up compelling routes for investigating granular lubrication.

Granular flows can be lubricated by adding smaller particles, either cohesionless hard grains or soft cohesive powders \cite{Wornyoh_Jasti_Fred_Higgs_2007, Madrid_Carlevaro_Pugnaloni_Kuperman_Bouzat_2021}.
Hard grains can act as ball bearings, forcing larger grains apart and reducing the net frictional force between grain surfaces, while grains of soft solid materials or hydrogels can act as compressible ball bearings or form a lubricating layer between surfaces \cite{Rudge_Sande_Dijksman_Scholten_2020}.
Adding self-repelling magnetic particles to neutral granular systems has been shown to help prevent clogging for flows passing through narrow apertures \cite{Nicolas_Ibáñez_Kuperman_Bouzat_2018}, but may not improve overall flow rates beyond what is expected for the addition of neutral small particles \cite{Carlevaro_Kuperman_Bouzat_Pugnaloni_Madrid_2022}.
Studies have also shown that near-frictionless soft particles experience clogging much more rarely than their hard, frictional counterparts \cite{Hong_Kohne_Morrell_Wang_Weeks_2017, Rudge_Sande_Dijksman_Scholten_2020}, suggesting that they may enhance flow.

Here, we discuss levitated granular materials with sound-induced attractive interactions that are easily tuned \textit{in situ}.
We explore the mechanical properties of granular rafts by performing shear tests, applying this approach to mesoscale samples composed of spheres or anisotropic sand grains.
We then add smaller spheres to these rafts to probe granular lubrication.

\section{Acoustic levitation}
When an ultrasound source is placed half a wavelength away from a reflector, a standing wave forms.
For a 40 kHz ultrasound wave in air, the wavelength is $\lambda \approx 8.5$ mm.
Objects can be levitated and trapped at the low-pressure zones at the wave nodes.
The acoustic radiation force which levitates particles is often referred to as the \textit{primary} force.
A conservative force, it can be computed from the gradient of the acoustic radiation potential, written as
\begin{align}\label{eq:gorkov}
    U_{\text{rad}} (\mb{r})
    =
    \frac{4 \pi a^3}{3} \bigg [ \frac{1}{2\rho_0 c_0^2} &f_0 \left \langle p(\mb{r}, t)^2 \right \rangle 
    \nonumber \\ 
    - \frac{3}{4} \rho_0 &f_1 \left \langle | \mb{v} (\mb{r}, t)|^2 \right \rangle \bigg ],
\end{align}
where $a$ is particle radius, $\rho_0$ and $c_0$ are the density and sound speed in air, respectively, $p(\mb{r}, t)$ and $\mb{v} (\mb{r}, t)$ are the pressure and velocity fields, and the angled brackets denote a time average \cite{Gorkov_1962, Lim_VanSaders_Jaeger_2024}.
The $f_i$ are scattering coefficients, defined as
\begin{equation}\label{eq:scatteringcoeff}
    f_0
    =
    1 - \frac{c_0^2 \rho_0}{c_p^2 \rho_p}
    ,~~
    f_1
    =
    \frac{2 (\rho_p / \rho_0 - 1)}{2 (\rho_p / \rho_0) + 1}
\end{equation}
where $\rho_p$ and $c_p$ are the density of the particle material and its sound speed.
From Eq.\ \ref{eq:scatteringcoeff} we find that for any solid the density contrast with air is sufficiently large that to very good approximation $f_0 \approx f_1 \approx 1$.
Under these conditions, the acoustic potential in Eq.\ \ref{eq:gorkov} is minimized where the mean square sound pressure $\left \langle p (\mb{r}, t)^2 \right \rangle = 0$, i.e., the primary acoustic force drives objects to levitate in a pressure nodal plane. 

When multiple objects are present in an acoustic cavity, sound scattered off their surfaces changes the pressure and velocity fields in Eq.\ \ref{eq:gorkov}, and this  results in \emph{secondary} interparticle forces.
These secondary scattering forces are anisotropic: they are attractive within the nodal plane but repulsive along the levitation axis.
For two spheres trapped in the nodal plane by the primary force, the  secondary scattering force within the levitation plane can be computed (neglecting viscous effects) as
\begin{equation}
    \mathbf{F}_{sc}
    \simeq
    - \frac{E_0 a^6}{r^4}, 
    \label{eq:silvabruus}
\end{equation}
where $r$ is the interparticle distance, the minus sign indicates that the force is attractive, and $E_0$ is the (constant) acoustic energy density \cite{Silva_Bruus_2014}.
This expression assumes we are in the Rayleigh limit where $a \ll \lambda$.
Note that the magnitude of the scattering force increases linearly with each particle's volume.
Since $E_0 \propto \left \langle p_{ac} \right \rangle^2$, where $p_{ac}$ is the acoustic pressure in the cavity, the strength of the  attraction between levitating particles can be tuned simply by changing the sound pressure.

The anisotropic nature of the secondary scattering forces in combination with their attractive nature, which applies to objects of various materials, shapes, and sizes, (see Eqs.\ \ref{eq:gorkov} and \ref{eq:scatteringcoeff}), results in the formation of quasi-two-dimensional particle rafts that float freely in the levitation plane.  
For monodisperse spheres, these rafts are comprised of close-packed crystalline regions \cite{Lim_VanSaders_Souslov_Jaeger_2022}.

In addition to the attractive interactions due to scattered sound, there can be sound-induced repulsive interaction between particles in the levitation plane.
This can occur when the particles' size is small and approaches a scale where the viscosity of air starts to play a significant role.
That scale is set by the thickness of the
viscous boundary layer $\delta_\nu = \sqrt{2 \nu / \omega}$ ($\nu$: kinematic viscosity, $\omega$: sound frequency), which forms around the surface of any stationary particle in an oscillating fluid (here, air).
The small but nonzero viscosity of air generates steady microstreaming flows around the particles, resulting in repulsive forces that can prevent particles from making contact.

In our experiments, $\delta_\nu \approx 11$ $\upmu$m, and viscous effects start playing a role for particles smaller than approximately 70 $\upmu$m.
For objects much larger than this, the attractive secondary scattering force dominates and and viscous microstreaming can be neglected.
However, for particles with sizes on the order of $\delta_\nu$, repulsive force starts to compete with the attraction from scattering.
Such particles in a raft no longer make direct contact, but instead sit at a nonzero interparticle distance from each other \cite{Wu_VanSaders_Lim_Jaeger_2023}.

For the same reason, sufficiently small particles also tend to sit at a distance from the surface of larger objects.
The attractive scattering force depends on the product of both objects' volumes (see Eq.\ \ref{eq:silvabruus}), so if one object is small enough, the repulsive microstreaming will still act to prevent the objects from coming into contact unless they are under significant stress.
Therefore, in mixtures of large and small grains, small particles below a certain size will be effectively frictionless, enabling their use in investigations of granular lubrication.

\section{Experimental setup}\label{sec:setup}
\begin{figure}[h]
\centering
\includegraphics[width=0.45\textwidth]{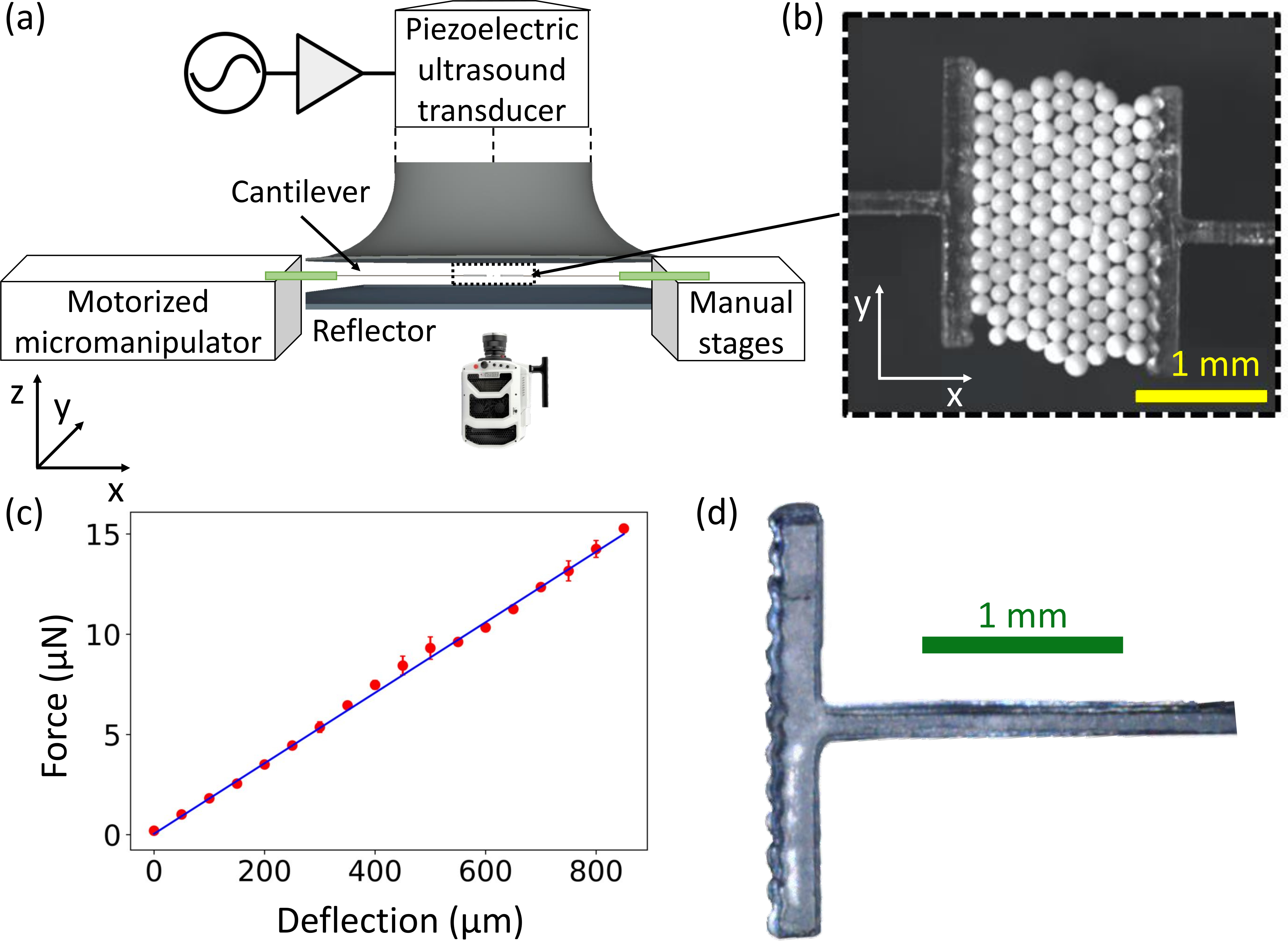}
\caption{
    Experimental setup and tools.
    (a) Schematic of the experimental system. 
    An AC signal is amplified to drive a piezoelectric ultrasound transducer connected to an aluminum horn. 
    The horn is placed such to create a standing wave between the horn and a transparent reflector.
    Experiments are filmed from below.
    Inside the acoustic cavity (dashed rectangle), granular particles are levitated and custom force sensors are used to perturb granular membranes and conduct measurements.
    (b) Experimental image corresponding to the dashed region in (a), viewed from below.
    Granular particles are levitated to form a raft and then trapped between two custom, T-shaped probe tips.
    Each probe tip is attached to the end of a cantilever.
    (c) Force vs.\ deflection curve of calibration data for one force sensor.
    (d) Close-up image of the probe tip used in (b), with surface features corresponding to the size of the spherical grains
}\label{fig:setup}
\end{figure}
We perform experiments with a single-axis acoustic levitation system, shown in Fig.\ \ref{fig:setup}(a).
A piezoelectric ultrasound transducer (Steiner \& Martins SMBLTD45F40H) is powered by a $f_0 = 40.45$ kHz sinusoidal signal from a function generator (BK Precision 4052) and a high-voltage amplifier (A.A. Lab Systems A-301 HS).
The transducer is attached to an aluminum horn to mechanically amplify the signal.
The end of the horn is positioned to create a half-wavelength standing wave between it and a clear glass reflector plate.
Granular particles are levitated in the acoustic cavity within this gap.
Here, we use 180-200 $\upmu$m high density polyethylene (HDPE) spheres, 40 $\upmu$m HDPE spheres, and anisotropic sand grains (Estes Ultra Stone aquarium sand, 100 $\upmu$m-1 mm).
A mirror angled at 45 degrees below the transparent reflector enables us to view and record levitating rafts  with a high-speed video camera (Phantom V2512) equipped with a zoom lens (Navitar Resolv4K).

Small ultrasound sensors (PUI Audio SMUT-1040K-TT) are mounted directly below the reflector.
The readings from these sensors track the acoustic pressure and are used for proportional-integral-derivative (PID) feedback control of the  pressure level in the cavity \cite{Brown_VanSaders_Kronenfeld_DeSimone_Jaeger_2024}.
Additionally, the temperature of the aluminum horn is controlled, as temperature changes can affect the horn's resonance properties.
The setup is enclosed in an acrylic box and placed on a passive vibration isolation platform (ThorLabs IsoPlate PTT900600) to further maintain a controlled environment.

We use small probes, which we insert from the sides into the cavity, to apply shear to levitated granular rafts and measure $\upmu$N-scale forces \cite{Brown_VanSaders_Kronenfeld_DeSimone_Jaeger_2024}.
A probe consists of a 3D-printed probe tip adhered to the end of a flexible cantilever.
Two of these probes are shown attached to a levitated granular raft in Fig.\ \ref{fig:setup}(b).
Each sensor is calibrated by pressing its tip on a $\upmu$g-resolution scale and gradually increasing the deflection. 
These weight measurements are then converted to forces.
The calibration data are fit to a line to obtain a calibration coefficient relating force to deflection.
An experimental calibration curve is shown in Fig.\ \ref{fig:setup}(c).
Force sensors are inserted to the levitation system from the side using a motorized micromanipulator (Eppendorf Patchman NP2) or a manual stage.
Forces are then computed from the positions of sensors in recorded images.

The probe tips are printed with high-resolution Continuous Liquid Interface Production (CLIP), a 3D printing technique \cite{Hsiao_Lee_Samuelsen_Lipkowitz_Kronenfeld_Ilyn_Shih_Dulay_Tate_Shaqfeh_et_al._2022, Lee_Hsiao_Lipkowitz_Samuelsen_Tate_DeSimone_2022, Kronenfeld_Rother_Saccone_Dulay_DeSimone_2024}.
CLIP uses a programmable micromirror array to project 2D images up through a transparent window to a vat of resin.
Ultraviolet light causes the resin to polymerize and harden.
Polymerization is inhibited in a `dead zone' layer, which prevents parts from sticking to the vat window.
A high-precision vertical stage moves the part upwards, enabling the fabrication of 3D structures.
This setup can result in a horizontal resolution of up to 2 $\upmu$m.
The probe tips used here were fabricated from a trimethylolpropane triacrylate (TMPTA)-based resin; a closeup image of one probe tip can be seen in Fig.\ \ref{fig:setup}(d).

The tips are T-shaped in order to confine the rafts and apply shear.
For rafts of spheres, the side of a tip facing the raft has rounded, particle-sized concavities.
This locally enhances the sound scattering intensity and thus promotes attraction of particles \cite{Lim_Jaeger_2023}.
The spacing of these features also encourages the formation of a single crystal with a particular orientation.
For experiments on anisotropic sand grains, bracket-shaped probe tips are used instead.

\section{Methods}\label{sec:methods}
\begin{figure}[h!]
\centering
\includegraphics[width=0.45\textwidth]{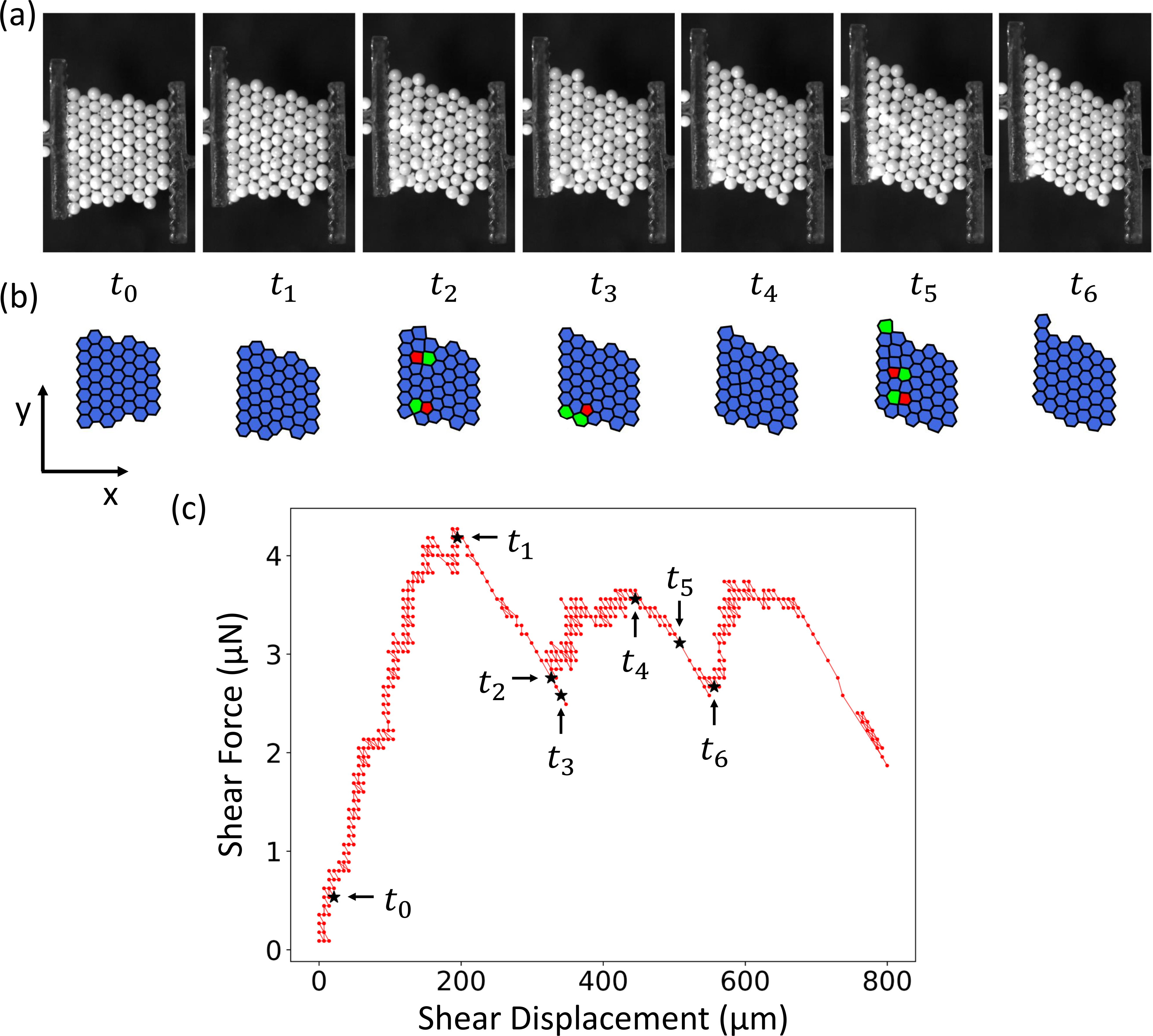}
\caption{
Experimentally shear testing an acoustically levitated granular raft composed of 180-200 $\upmu$m spheres.
(a) Time series of images from throughout a shear test, viewed from below.
The probe on the left moves in the $+y$ direction to shear the raft, while the probe on the right measures the shear force.
To better visualize the relative displacement between the probes the images in this sequence have been shifted vertically so that the right probe appears to remain at the same position.
(b) Voronoi diagrams corresponding to each image in (a). Particles are colored by number of nearest neighbors nearest neighbors (blue: 6, red: 5, green: 7), with edge particles excluded from the diagram.
(c) Plot of shear force vs.\ displacement for this experiment.
Data is shown in red, while the times of the images in (a) are labeled as black stars.
The raft begins in a crystalline state. 
The shear force increases to a peak, before the raft experiences several successive slip events during which it deforms plastically
}
\label{fig:shear_ex}
\end{figure}

By moving the probe tips under computer control, we can perform mechanical shear tests on acoustically levitated granular rafts.
This is pictured in Fig.\ \ref{fig:shear_ex}.
These tests mimic traditional mechanical testing performed on macroscale samples of materials such as metals, but also bear resemblance to granular shear testing on geophysical materials like soil.

Once a sample of particles has been levitated and has self-assembled into a floating raft, probes are attached on either side, using the sound-induced secondary scattering force to provide adhesion. The probes are then  moved to shear the material to the point of plastic deformation.
Other types of mechanical testing are also possible, though they may introduce complications: in tensile testing sample delamination from the probe tips can occur, and compression testing can cause the quasi-two-dimensional sample to buckle out of plane.
Shear testing also keeps the sample relatively close to the center of the acoustic cavity where the potential is strongest \cite{Brown_VanSaders_Kronenfeld_DeSimone_Jaeger_2024}.

Prior to performing experiments, the electronics and temperature control system are turned on and allowed to run for at least 30 minutes to ensure consistent conditions.
The gap between the horn and reflector is adjusted to maximize the output acoustic pressure near the cavity, as measured by the sensors placed below the reflector.
Granular particles are inserted into the center of the cavity using a mesh spatula, where they levitate and rapidly form a close-packed two-dimensional raft.
Two probes are inserted into the cavity and moved along the $x$-direction (see Figs.\ \ref{fig:setup}(a) and \ref{fig:shear_ex}(a) for the alignment of the coordinate system) close to the raft.
Because the presence of objects in a resonance cavity can affect its resonance characteristics, the horn-reflector gap is again adjusted to maximize acoustic pressure.
The acoustic pressure control is switched on to maintain the cavity energy density at a constant level, adjusting for slight changes in environmental conditions (e.g., room temperature).
The probes are then moved to make contact with the raft's edges.
For rafts comprised of spheres, we furthermore generate an initial, defect-free configuration by applying small amplitude cyclic shear in the $\pm y$ directions prior to commencing a large-strain shear test.

\section{Granular rafts of spheres}\label{sec:spheres}
We performed 110 shear experiments at the same acoustic pressure value ($p_{ac} = 1040$ Pa) on rafts of various sizes, ranging from 8 to 215 spheres.
Shear stress and strain were calculated from measured force, displacement, number of particles, and raft width, given that each raft was approximately rectangular.
An exemplary shear test, on a granular raft containing 107 spheres of diameter 180-200 $\upmu$m, is visualized in Fig.\  \ref{fig:shear_ex}.
The probe on the right is attached to a cantilever that deflects in $\pm y$ to measure shear force. 
The probe on the left is attached to a stiffer cantilever that deflects in $\pm x$.
The left probe is moved in the $+y$ direction at a constant rate of 20 $\upmu$m/s to apply shear strain.
Figure \ref{fig:shear_ex}(a) shows a time series of images from one such experiment, and (b) shows Voronoi diagrams for each of these frames, colored by each particle's number of nearest neighbors (blue: 6, red: 5, green: 7).

At time $t_0$, the sample is in the initial, crystalline state.
As strain is applied, the shear force increases to a maximum near $t_1$,  while the packing maintains essentially the same configuration of particles and only small fluctuations in stress occur as particles ride up on each other.
This changes once the shear displacement has reached a value corresponding to a particle diameter and the first significant slip event is initiated at $t_1$.
Between $t_1$ and $t_2$, defects form and propagate as particles slip past each other; $t_2$ shows two dislocation pairs, each consisting of one five-neighbored particle adjacent to a seven-neighbored particle.
This slip also corresponds to a large drop in shear force.
At $t_3$, the raft has taken on a new, nearly crystalline state (with some disorder near the raft's edge).
The shear force increases again to $t_4$, when another slip event occurs from $t_4$ to $t_6$.
Dislocation pairs are visible at $t_5$.
By $t_6$, the slip event has ended and the raft has taken on a new crystalline configuration.
Shear force vs.\ displacement data is plotted in Fig.\ \ref{fig:shear_ex}(c). 
This repeating process of  stress build-up followed by stress relaxation through a  slip event is commonly observed in granular materials \cite{Krim_Yu_Behringer_2011} and small samples of atomic materials \cite{Kim_Greer_2009}.
In granular bulk media, this process  is taken as the basis for the critical state, where, under fixed normal load,  the average shear stress reaches a level that is independent of applied strain.
With the probe setup in Fig.\ \ref{fig:shear_ex} we can now initiate and measure the underlying individual, local deformation and slip events in a controlled manner.

For each shear experiment, we compute the yield shear stress and strain at the first major slip event, as well as an effective shear modulus from the initial slope of the stress-strain curves.
The time of the first slip event is identified manually.
The shear modulus is computed by fitting a line to the stress vs.\ strain data during the initial stress increase.
For some experiments, the first slip event was not a sharp peak and instead slightly rounded.
For this reason, only the first half of the data in this region were used for computing the effective shear modulus.

\begin{figure*}[h!]%
\centering
\includegraphics[width=0.95\textwidth]{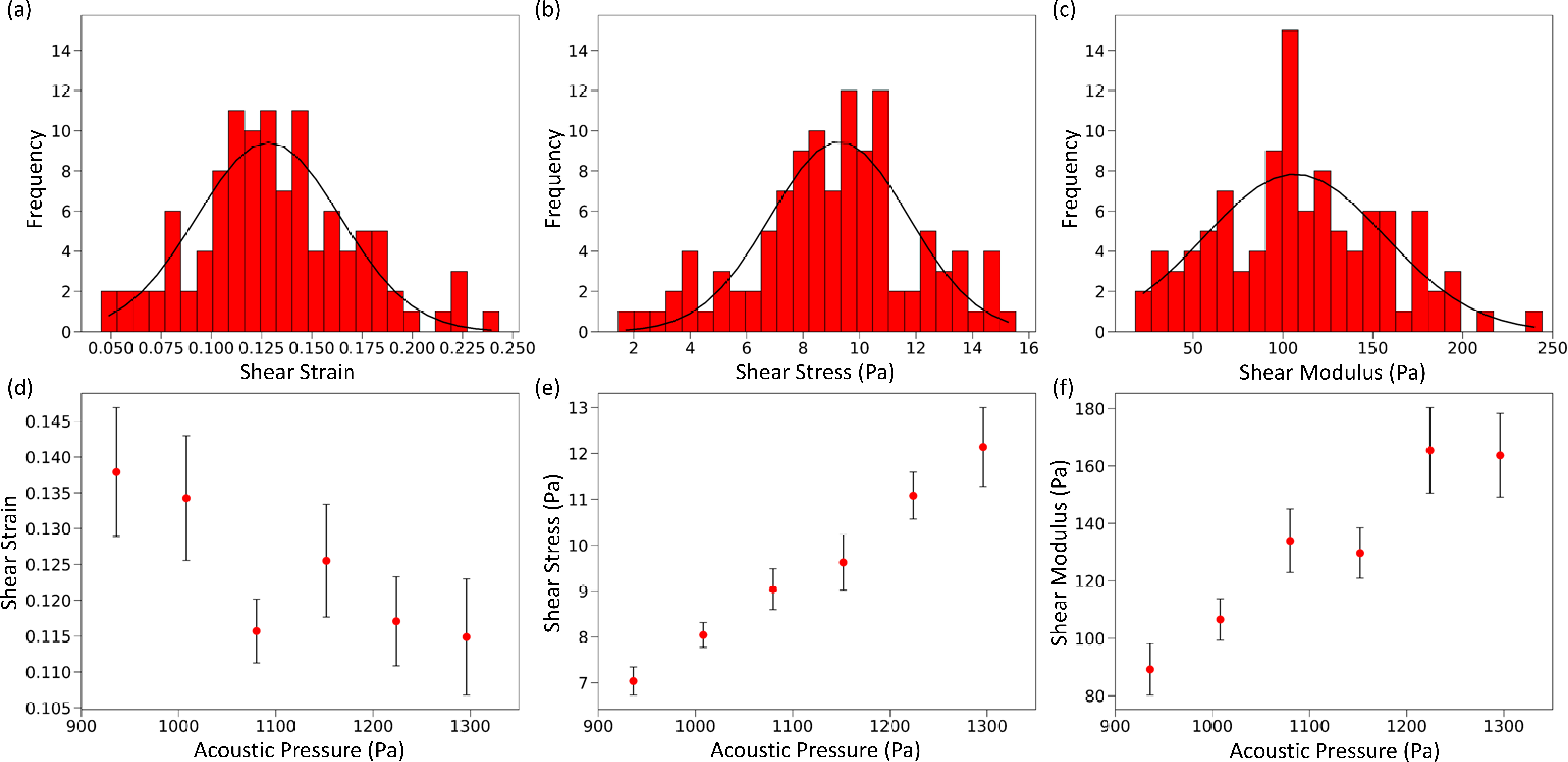}
\caption{
Results from shear experiments on acoustically levitated granular rafts of spherical particles, conducted at the same measured acoustic pressure.
(a-c) Histograms for an ensemble of shear tests conducted at the same acoustic pressure with Gaussian distributions (black) fit to the data.
(a) Yield shear strain at the first slip event.
(b) Yield shear stress at the first slip event.
(c) Effective shear modulus during initial deformation.
(d-f) Data for shear tests conducted at different acoustic pressure values.
(d) Yield shear strain vs.\ acoustic pressure.
(e) Yield shear stress vs.\ acoustic pressure.
(f) Effective shear modulus vs.\ acoustic pressure.
}\label{fig:sphereshear}
\end{figure*}

Using data from all 110 shear experiments, we can compile histograms of the yield strain, yield stress and the effective shear modulus. 
These  are plotted in Fig.\ \ref{fig:sphereshear}(a), (b), and (c) respectively, with Gaussian fits to the distributions.
Typically, the first significant slip events occurred at strain values $\gamma = 0.129 \pm 0.003$ (mean $\pm$ standard error for an ensemble of 110 shear tests) and stress values $\sigma = 9.3 \pm 0.2$ Pa.
The levitating granular rafts had shear moduli $ G = 107 \pm 5$ Pa.
This very soft behavior is a result of $\upmu$N-scale interparticle forces and the quasi-two-dimensional nature of the free-floating structures.

While the mean of these distributions is well defined, their  considerable width is a reflection of the small raft size and the extreme sensitivity of granular matter to details of the local contacts. 
With so few particles in each raft, a raft's deformation and yielding behavior will depend strongly on minute variances in the particle shape and (frictional) surface properties.
Thus, despite the fact that all particles are `spheres' of the same material and of  similar size assembled into a close-packed, highly ordered configuration, there can be significant variances as far as the contact forces are concerned.
As such, the data in Figs. \ref{fig:shear_ex}(c) and \ref{fig:sphereshear}(a-c) more generally reflect the meso-scale, few-grain behavior expected also from other granular media, effectively providing a zoomed-in view of deformations inside a shearing zone.   

We did not observe a statistically meaningful trend with the number of spheres in rafts ranging from 8 to 215 particles.
This may be a consequence of the significant variance around the mean rather than a reliable indication that no such trend could be present.
The lack of a clear trend differs from what has been observed for mesoscopic crystals of  atoms. 
Nanocrystal metal pillars have been found to increase their strength with decreasing size \cite{Kim_Greer_2009, Greer_Nix_2006, Brinckmann_Kim_Greer_2008}.
This strengthening is derived from dislocation starvation, where plasticity requires the formation of new defects rather than just the motion of existing ones. 
This phenomenon could potentially apply to acoustically-bound granular crystals, but compared to the atomic systems our rafts appear to be already in the ultra-small, starved limit.  
A thorough investigation and comparison would require the ability to levitate significantly larger granular rafts of a few thousand particles.
However, in our current experimental setup the maximum raft size is limited to approximately 250 particles, which is dictated by the requirement to fit within the central region of the acoustic cavity in which the primary force has relatively uniform strength \cite{Brown_VanSaders_Kronenfeld_DeSimone_Jaeger_2024}.

Changing the sound pressure $p_{ac}$ in the cavity and thereby the acoustic energy density $E_0$ in Eq. \ref{eq:silvabruus} by adjusting the amplitude of the electrical signal that drives the ultrasound transducer also does not change, in a statistically significant manner, the width of the distributions in yield strain, yield stress, and effective modulus.
However, it provides a straightforward way to tune the mean value of these quantities.  
Data from an additional 102 experiments for different rafts levitated in fields with acoustic pressure values from 940 - 1300 Pa are shown in Fig.\ \ref{fig:sphereshear}(d-f).
We find that the mean yield strain decreases with acoustic pressure while both the mean yield shear stress and the mean effective modulus increase.
Specifically, while the yield strain changes by less than 20\% and thus stays well within the original distribution, the yield stress and effective modulus can be varied by a factor of about 2.
Increasing acoustic pressure strengthens the attractive interparticle forces that bind particles together, stiffening the raft.
Thus, the ability to change the acoustic pressure in the cavity at will allows for easy control over a granular raft's mechanical properties.

\section{Anisotropic particles and lubrication}\label{sec:sand}
\begin{figure}[h!]
    \centering
    \includegraphics[width=0.43\textwidth]{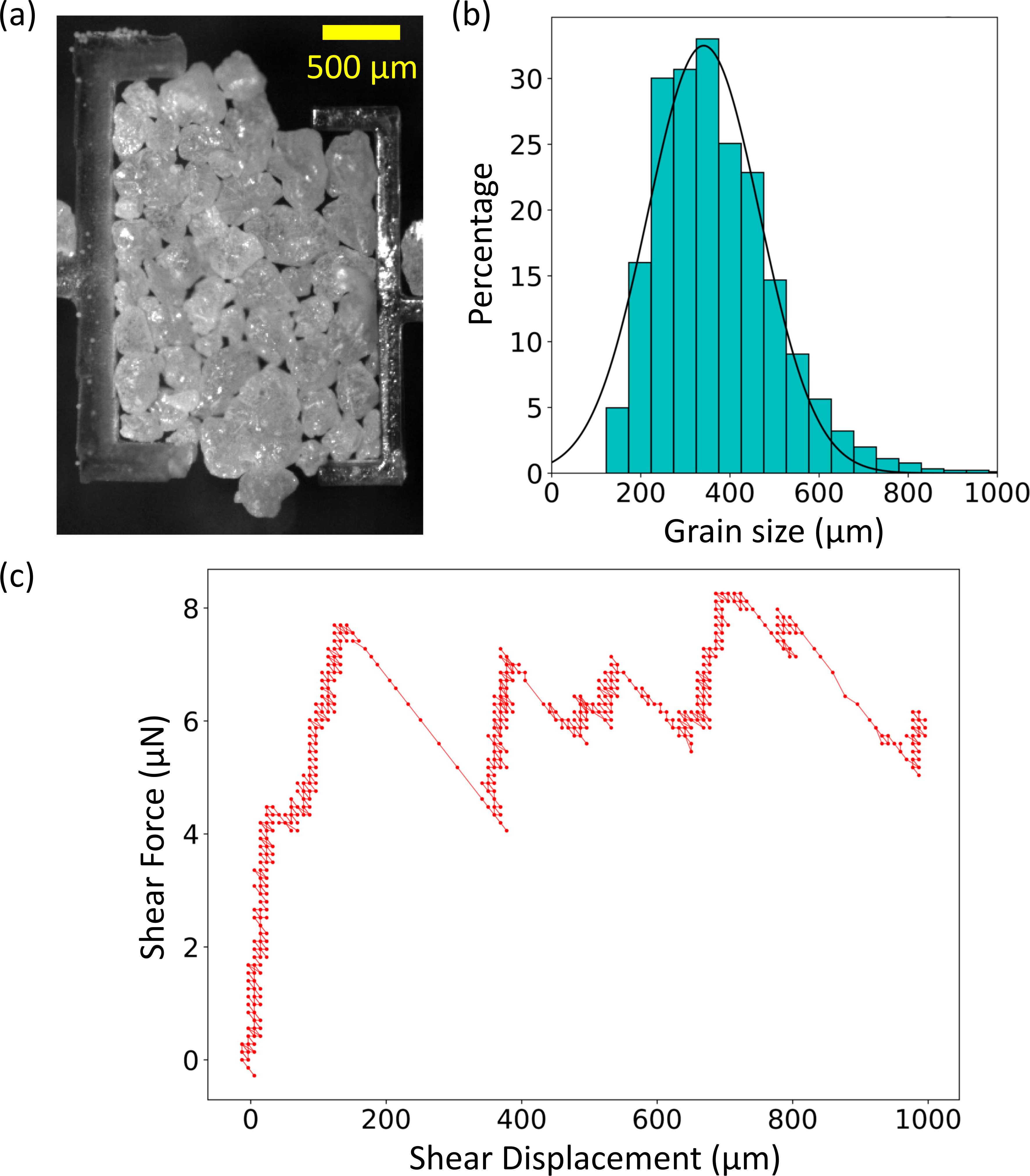}
    \caption{
    Acoustically levitated rafts composed of sand grains.
    (a) A raft of sand grains trapped between two bracket-shaped probes ($\phi = 0.954$).
    (b) Histogram of sand grain sizes for a representative sample.
    (c) Plot of shear force vs.\ displacement for one experiment on a raft composed of sand grains.
    }
    \label{fig:shear_ex_sand}
\end{figure}

We performed similar mechanical shear tests on rafts composed of anisotropic  silica sand grains, which are irregular in shape and have a rough surface texture.
Like the spherical particles, levitated sand grains pack closely to form a quasi-two-dimensional raft; one such raft can be seen in Fig.\ \ref{fig:shear_ex_sand}(a).
The sand grains used have size $300 \pm 100$ $\upmu$m (mean $\pm$ standard deviation), eccentricity $0.7 \pm 0.1$, convexity $0.98 \pm 0.02$, and Wadell roundness $0.37 \pm 0.07$. 
These values were computed for representative samples of at least 150 grains and each distribution fit to a Gaussian (e.g., grain size in Fig.\ \ref{fig:shear_ex_sand}(b)). 
The shear experiments on these rafts used force probes with bracket-shaped tips, shown in Fig.\ \ref{fig:shear_ex_sand}(a).

A shear force vs.\ displacement plot from one such experiment is shown in Fig.\ \ref{fig:shear_ex_sand}(c).
The analogous curve for a raft of spherical particles (Fig.\ \ref{fig:shear_ex}(c)) shows slip events that are relatively equally spaced, corresponding to the size of the spheres.
Here, the irregularity of the sand grains results in a more erratic curve.
The raft in Fig.\ \ref{fig:shear_ex_sand}(a) was levitated in the acoustic trap and sheared and then brought back to zero displacement repeatedly for relatively small strains (typically without large slip events) at an acoustic pressure of 1080 Pa.
The average effective shear modulus for these rafts composed of sand grains was $G = 180 \pm 10$ Pa (mean $\pm$ standard error).
This modulus is $\approx$25\% larger than what we would expect for a crystalline raft of 180-200 $\upmu$m spheres at a similar acoustic pressure (Fig.\ \ref{fig:sphereshear}(f)), reflecting how the combination of larger average grain size, the corresponding increase in interparticle attraction, and the shape anisotropy of the sand grains results in more resistance to shear.

\begin{figure}[h!]
    \centering
    \includegraphics[width=0.45\textwidth]{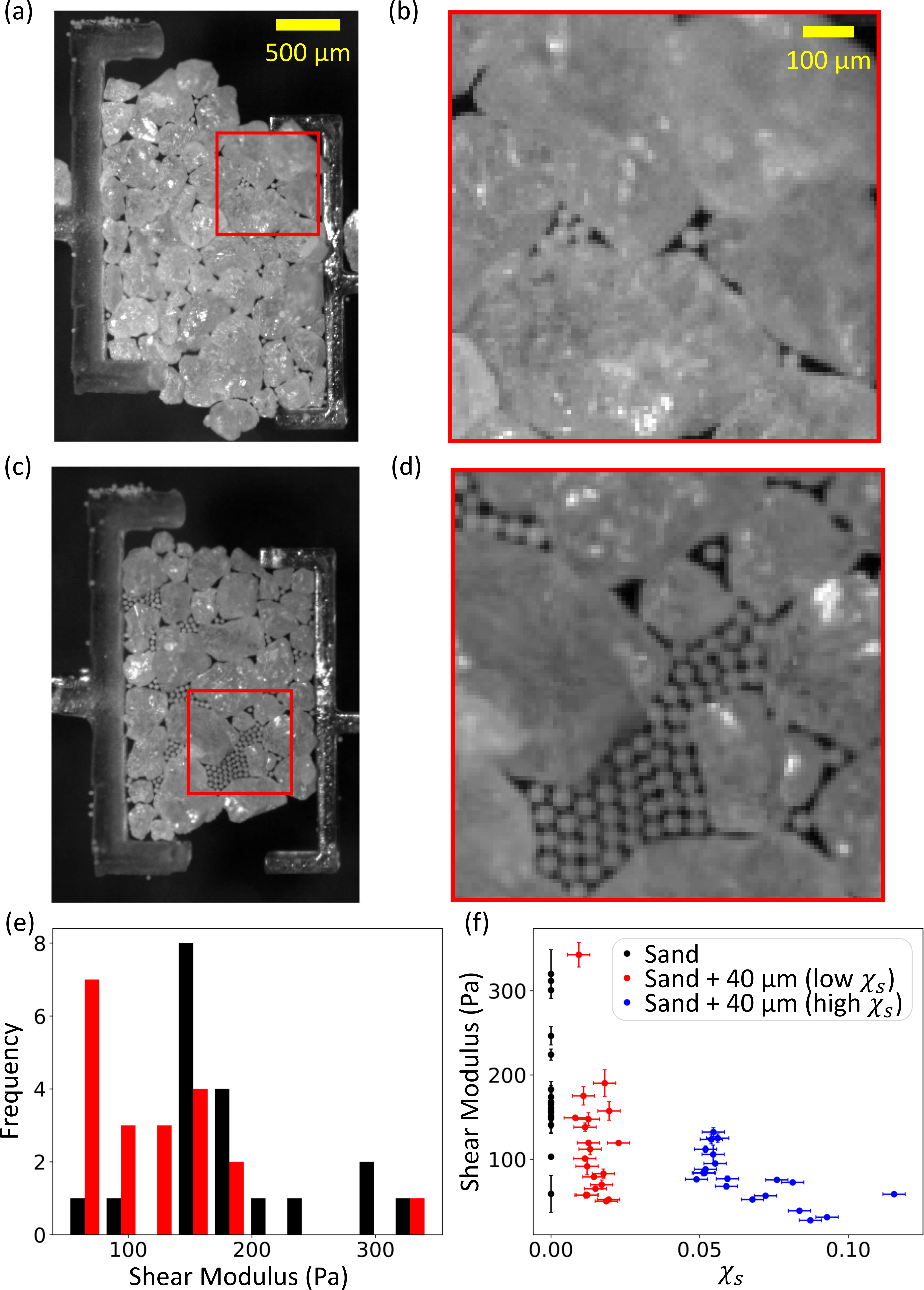}
    \caption{
    Shear experiments on acoustically levitated rafts of sand mixed with 40 $\upmu$m particles.
    (a) A raft of sand grains with a small concentration of 40 \um spheres ($\phi = 0.963$, $\chi_s = 0.011$).
    (b) A zoomed-in image corresponding to the red box in (a).
    (c) A raft of sand grains with a higher concentration of 40 \um spheres ($\phi = 0.935$, $\chi_s = 0.072$).
    (d) A zoomed-in image corresponding to the red box in (c).
    (e) Histogram of effective shear modulus for sand (black) and sand with small spheres (red).
    (f) Effective shear modulus vs.\ small particle concentration $\chi_s$ for rafts of only sand (black), sand with small concentrations of 40 \um particles (red), and sand with higher concentrations of 40 \um particles (blue)
    }
    \label{fig:shear_sand_with40}
\end{figure}

When 40 $\upmu$m `fines' (HDPE spheres) are added to a raft of sand particles, they automatically assemble into the interstitial spaces between larger grains.
This can be seen in Fig.\ \ref{fig:shear_sand_with40}(a-d).
Because microstreaming in the viscous boundary layer repels these smaller particles from each other and the sand grains, they do not touch, and there is effectively no friction between the two species of particles.
Sand rafts with and without 40 $\upmu$m particles have different area fractions $\phi$ of solids and small particle concentrations $\chi_s$.
We define the overall area fraction as 
\begin{equation}\label{eq:areafrac}
    \phi
    =
    \frac{A_s + A_L}{A}
\end{equation}
where $A_L$ is the area occupied by the large sand grains, $A_s$ is the area occupied by the 40 $\upmu$m spheres, and $A$ is the total area of the raft.
We then define the small particle concentration as
\begin{equation}
    \chi_s
    =
    \frac{A_s}{A_s + A_L}.
\end{equation}

A raft composed only of sand has $\chi_s = 0$, and likely has some empty gaps between particles due to the anisotropy of the grains.
If small spheres are added to this raft, they can partially fill the gaps, so $\chi_s$ and $\phi$ increase. 
While Fig.\ \ref{fig:shear_ex_sand}(a)  shows a  raft with only sand grains, Fig.\ \ref{fig:shear_sand_with40}(a, b) show the same raft following the addition of small spheres.
In this case, $\phi$ increased by 0.009 with a $\chi_s$ of 0.011.
As more small spheres are added, they tend to `pool.'
This is visible in Fig.\ \ref{fig:shear_sand_with40}(c, d), which show a mixed raft with $\phi = 0.935$ and $\chi_s = 0.072$.
The relationship between $\phi$ and $\chi_s$ for closely-packed rafts is non-monotonic, similar to packings of bidisperse spheres \cite{Anzivino_Casiulis_Zhang_Moussa_Martiniani_Zaccone_2023}.
A densely packed raft composed only of large particles would have a higher $\phi$ than a mixed raft with large gaps filled by small spheres.

To see how this affects the response to shear, small concentrations of 40 $\upmu$m particles were added to the same sand raft as in Fig.\ \ref{fig:shear_ex_sand}(a), and more shear tests were performed.
Images of one of these experiments are shown in Fig.\ \ref{fig:shear_sand_with40}(a, b).
The measured effective shear moduli for these experiments are shown in Fig.\ \ref{fig:shear_sand_with40}(e), with data for the raft both without (black) and with (red) 40 $\upmu$m spheres.
The average shear modulus for these levitated rafts composed of a mixture of sand and a low concentration of small particles was $G = 120 \pm 10$ Pa.
The considerable scatter in the data is, as with the rafts in Sec.\ \ref{sec:spheres}, a result of the mesoscale size of the system.
Here, the sand-only rafts had a mean area fraction of $\phi = 0.957$ and the addition of the 40 \um particles at a concentration of less than $\chi_s = 0.02$, or 2\%, increased the area fraction by approximately 0.006.
At the same time, this was enough to result in a decrease in shear modulus of more than 30\% from the sand-only raft.

We next performed experiments on a different set of rafts with higher concentrations  $\chi_s$ of 40 $\upmu$m spheres (e.g., Fig.\ \ref{fig:shear_sand_with40}(c, d)). 
For these rafts, the area fraction $\phi$ was lower than for the low $\chi_s$ rafts, as having significantly-sized regions of small spheres sitting at finite interparticle distance mandates additional free space within the raft.
In this set of experiments, the number of small particles present decreased through different trials as particles were forced out by displacements of the sand grains or excited by local increases in acoustic pressure, allowing us to examine how $\chi_s$ affects a raft's shear modulus.
Figure \ref{fig:shear_sand_with40}(f) plots shear modulus vs.\ $\chi_s$ for the raft of just sand (black), the raft with small $\chi_s$ (red), and then rafts with higher $\chi_s$ (blue).
As $\chi_s$ increases, the shear modulus value decreases significantly.

To ascertain the contribution   to this shear resistance from relative movement among the small particles, we also measured the effective shear modulus of a raft composed solely of the 40 $\upmu$m particles.
This modulus was found to be $G = 1.0 \pm 0.3$ Pa, two orders of magnitude smaller than the shear modulus for the rafts containing larger grains.
Thus, the 40 $\upmu$m particles  themselves provide only minimal resistance to shear.

We can rationalize the softening of the granular composite material when 40 $\upmu$m spheres are added by considering these small particles as having a hard-sphere core embedded in a soft shell. 
The $\delta_\nu \approx 11$ $\upmu$m viscous boundary layer that surrounds each of these spheres creates a lubricating layer of air that surrounds each HDPE core, acting to prevent frictional contacts.
Under stress, the effective radius $a+\delta_{\nu}$ of these core-shell particles can be compressed slightly.
This produces behavior similar to soft, frictionless spheres in other granular systems, which have previously been found to exhibit lubricating effects \cite{Rudge_Sande_Dijksman_Scholten_2020, Hong_Kohne_Morrell_Wang_Weeks_2017}.
Conversely, if these small particles are lost from a raft, some voids are left.
Without the small particles and their viscous boundary layers forcing the sand grains apart, some of those voids close and increase the contact between sand grains, thus increasing the overall friction within the granular raft and stiffening the sample.

Therefore, our results with acoustically bound particles show that the lubricating effects arising from soft, frictionless particles are also applicable to systems with cohesive grains,
which typically exhibit increased clogging and slower flow compared to their non-cohesive  counterparts \cite{Mandal_Nicolas_Pouliquen_2020, Rognon_Roux_Naaïm_Chevoir_2008}.

\section{Conclusion}
Granular particles levitated in an acoustic field form a quasi-two-dimensional material with cohesive properties that can be varied \textit{in situ}.
These levitated rafts float freely, allowing for closer investigation of their innate characteristics without the effects of the fixed boundary conditions imposed by a container.
The interparticle attraction from acoustic scattering does not require particle-particle contact and is able to act at a distance to bring particles together.
The interparticle cohesion can be changed between experiments, even for the same sample, allowing for better isolation of the effects of changing interaction strength.
Acoustic levitation, along with the experimental methods described here to measure and apply strain and stress, therefore allows for unique investigations into the physics of granular materials.

In particular, the presence of both attractive scattering and  repulsive microstreaming forces provides access to multi-phase regimes where grains can interact via cohesive contacts as well as via granular lubrication, depending on their size. 
We demonstrate that adding effectively frictionless fines ($\lesssim 70$ $\upmu$m) to a sample of larger, attractive sand grains decreases its shear modulus, and that the change in shear modulus depends on the concentration of lubricant particles.
The lubricant particles sit at finite interparticle distance; under stress, they can be compressed closer together, acting like soft spheres, but typically do not make contact with each other or with the sand grains.
As a result, the addition of soft, frictionless particles to a granular material (even one made of highly anisotropic and attractive or cohesive grains) tends to decrease its stiffness despite increasing the overall packing fraction.

\backmatter

%\bmhead{Supplementary information}

\bmhead{Acknowledgements}
We thank Brady Wu, Qinghao Mao, and Tali Khain for many useful discussions.
This research was supported by the National Science Foundation through grant DMR-2104733.
We utilized shared equipment and resources at the University of Chicago MRSEC, which is supported by the National Science Foundation through grant DMR-2011854.
Additional support was provided by the Army Research Office through award W911NF-21-2-0146 (HMJ).
JMK acknowledges support through the National Science Foundation Graduate Research Fellowship Program under grant DGE-1656518.

\bibliography{draft}% common bib file

\begin{thebibliography}{10}
\expandafter\ifx\csname url\endcsname\relax
  \def\url#1{\burl{#1}}\fi
\expandafter\ifx\csname urlprefix\endcsname\relax\def\urlprefix{URL }\fi
\providecommand{\bibinfo}[2]{#2}
\providecommand{\eprint}[2][]{\url{#2}}
\providecommand{\doi}[1]{\url{https://doi.org/#1}}
\bibcommenthead

\bibitem{Ralaiarisoa_Dupont_Moctar_Naaim-Bouvet_Oger_Valance_2022}
\bibinfo{author}{Ralaiarisoa, V.} \emph{et~al.}
\newblock \bibinfo{title}{Particle impact on a cohesive granular media}.
\newblock \emph{\bibinfo{journal}{Physical Review E}}
  \textbf{\bibinfo{volume}{105}}, \bibinfo{pages}{054902}
  (\bibinfo{year}{2022}).

\bibitem{Castellanos_2005}
\bibinfo{author}{Castellanos, A.}
\newblock \bibinfo{title}{The relationship between attractive interparticle
  forces and bulk behaviour in dry and uncharged fine powders}.
\newblock \emph{\bibinfo{journal}{Advances in Physics}}
  \textbf{\bibinfo{volume}{54}}, \bibinfo{pages}{263–376}
  (\bibinfo{year}{2005}).

\bibitem{Kolehmainen_Ozel_Gu_Shinbrot_Sundaresan_2018}
\bibinfo{author}{Kolehmainen, J.}, \bibinfo{author}{Ozel, A.},
  \bibinfo{author}{Gu, Y.}, \bibinfo{author}{Shinbrot, T.} \&
  \bibinfo{author}{Sundaresan, S.}
\newblock \bibinfo{title}{Effects of polarization on particle-laden flows}.
\newblock \emph{\bibinfo{journal}{Physical Review Letters}}
  \textbf{\bibinfo{volume}{121}}, \bibinfo{pages}{124503}
  (\bibinfo{year}{2018}).

\bibitem{Sánchez_Scheeres_2020}
\bibinfo{author}{Sánchez, P.} \& \bibinfo{author}{Scheeres, D.~J.}
\newblock \bibinfo{title}{Cohesive regolith on fast rotating asteroids}.
\newblock \emph{\bibinfo{journal}{Icarus}} \textbf{\bibinfo{volume}{338}},
  \bibinfo{pages}{113443} (\bibinfo{year}{2020}).

\bibitem{Hemmerle_Schröter_Goehring_2016}
\bibinfo{author}{Hemmerle, A.}, \bibinfo{author}{Schröter, M.} \&
  \bibinfo{author}{Goehring, L.}
\newblock \bibinfo{title}{A cohesive granular material with tunable
  elasticity}.
\newblock \emph{\bibinfo{journal}{Scientific Reports}}
  \textbf{\bibinfo{volume}{6}}, \bibinfo{pages}{35650} (\bibinfo{year}{2016}).

\bibitem{Gans_Pouliquen_Nicolas_2020}
\bibinfo{author}{Gans, A.}, \bibinfo{author}{Pouliquen, O.} \&
  \bibinfo{author}{Nicolas, M.}
\newblock \bibinfo{title}{Cohesion-controlled granular material}.
\newblock \emph{\bibinfo{journal}{Physical Review E}}
  \textbf{\bibinfo{volume}{101}}, \bibinfo{pages}{032904}
  (\bibinfo{year}{2020}).

\bibitem{Peters_Lemaire_2004}
\bibinfo{author}{Peters, F.} \& \bibinfo{author}{Lemaire, E.}
\newblock \bibinfo{title}{Cohesion induced by a rotating magnetic field in a
  granular material}.
\newblock \emph{\bibinfo{journal}{Physical Review E}}
  \textbf{\bibinfo{volume}{69}}, \bibinfo{pages}{061302}
  (\bibinfo{year}{2004}).

\bibitem{Lehman_Christman_Jacobs_Johnson_Palchoudhuri_Tieman_Vajpeyi_Wainwright_Walker_Wilson_et_al._2022}
\bibinfo{author}{Lehman, S.~Y.} \emph{et~al.}
\newblock \bibinfo{title}{Universal aspects of cohesion}.
\newblock \emph{\bibinfo{journal}{Granular Matter}}
  \textbf{\bibinfo{volume}{24}}, \bibinfo{pages}{35} (\bibinfo{year}{2022}).

\bibitem{Lumay_Vandewalle_2007}
\bibinfo{author}{Lumay, G.} \& \bibinfo{author}{Vandewalle, N.}
\newblock \bibinfo{title}{Tunable random packings}.
\newblock \emph{\bibinfo{journal}{New Journal of Physics}}
  \textbf{\bibinfo{volume}{9}}, \bibinfo{pages}{406} (\bibinfo{year}{2007}).

\bibitem{Lumay_Vandewalle_2008}
\bibinfo{author}{Lumay, G.} \& \bibinfo{author}{Vandewalle, N.}
\newblock \bibinfo{title}{Controlled flow of smart powders}.
\newblock \emph{\bibinfo{journal}{Physical Review E}}
  \textbf{\bibinfo{volume}{78}}, \bibinfo{pages}{061302}
  (\bibinfo{year}{2008}).

\bibitem{Brown_VanSaders_Kronenfeld_DeSimone_Jaeger_2024}
\bibinfo{author}{Brown, N.~M.}, \bibinfo{author}{VanSaders, B.},
  \bibinfo{author}{Kronenfeld, J.~M.}, \bibinfo{author}{DeSimone, J.~M.} \&
  \bibinfo{author}{Jaeger, H.~M.}
\newblock \bibinfo{title}{Direct measurement of forces in air-based acoustic
  levitation systems}.
\newblock \emph{\bibinfo{journal}{Review of Scientific Instruments}}
  \textbf{\bibinfo{volume}{95}}, \bibinfo{pages}{094901}
  (\bibinfo{year}{2024}).

\bibitem{Lim_VanSaders_Jaeger_2024}
\bibinfo{author}{Lim, M.~X.}, \bibinfo{author}{VanSaders, B.} \&
  \bibinfo{author}{Jaeger, H.~M.}
\newblock \bibinfo{title}{Acoustic manipulation of multi-body structures and
  dynamics}.
\newblock \emph{\bibinfo{journal}{Reports on Progress in Physics}}
  \textbf{\bibinfo{volume}{87}}, \bibinfo{pages}{064601}
  (\bibinfo{year}{2024}).

\bibitem{Wornyoh_Jasti_Fred_Higgs_2007}
\bibinfo{author}{Wornyoh, E. Y.~A.}, \bibinfo{author}{Jasti, V.~K.} \&
  \bibinfo{author}{Fred~Higgs, I., C.}
\newblock \bibinfo{title}{A review of dry particulate lubrication: Powder and
  granular materials}.
\newblock \emph{\bibinfo{journal}{Journal of Tribology}}
  \textbf{\bibinfo{volume}{129}}, \bibinfo{pages}{438–449}
  (\bibinfo{year}{2007}).

\bibitem{Madrid_Carlevaro_Pugnaloni_Kuperman_Bouzat_2021}
\bibinfo{author}{Madrid, M.~A.}, \bibinfo{author}{Carlevaro, C.~M.},
  \bibinfo{author}{Pugnaloni, L.~A.}, \bibinfo{author}{Kuperman, M.} \&
  \bibinfo{author}{Bouzat, S.}
\newblock \bibinfo{title}{Enhancement of the flow of vibrated grains through
  narrow apertures by addition of small particles}.
\newblock \emph{\bibinfo{journal}{Physical Review E}}
  \textbf{\bibinfo{volume}{103}}, \bibinfo{pages}{L030901}
  (\bibinfo{year}{2021}).

\bibitem{Rudge_Sande_Dijksman_Scholten_2020}
\bibinfo{author}{D.~Rudge, R.~E.}, \bibinfo{author}{Sande, J. P. M. v.~d.},
  \bibinfo{author}{A.~Dijksman, J.} \& \bibinfo{author}{Scholten, E.}
\newblock \bibinfo{title}{Uncovering friction dynamics using hydrogel particles
  as soft ball bearings}.
\newblock \emph{\bibinfo{journal}{Soft Matter}} \textbf{\bibinfo{volume}{16}},
  \bibinfo{pages}{3821–3831} (\bibinfo{year}{2020}).

\bibitem{Nicolas_Ibáñez_Kuperman_Bouzat_2018}
\bibinfo{author}{Nicolas, A.}, \bibinfo{author}{Ibáñez, S.},
  \bibinfo{author}{Kuperman, M.~N.} \& \bibinfo{author}{Bouzat, S.}
\newblock \bibinfo{title}{A counterintuitive way to speed up pedestrian and
  granular bottleneck flows prone to clogging: can ‘more’ escape faster?}
\newblock \emph{\bibinfo{journal}{Journal of Statistical Mechanics: Theory and
  Experiment}} \textbf{\bibinfo{volume}{2018}}, \bibinfo{pages}{083403}
  (\bibinfo{year}{2018}).

\bibitem{Carlevaro_Kuperman_Bouzat_Pugnaloni_Madrid_2022}
\bibinfo{author}{Carlevaro, C.~M.}, \bibinfo{author}{Kuperman, M.~N.},
  \bibinfo{author}{Bouzat, S.}, \bibinfo{author}{Pugnaloni, L.~A.} \&
  \bibinfo{author}{Madrid, M.~A.}
\newblock \bibinfo{title}{On the use of magnetic particles to enhance the flow
  of vibrated grains through narrow apertures}.
\newblock \emph{\bibinfo{journal}{Granular Matter}}
  \textbf{\bibinfo{volume}{24}}, \bibinfo{pages}{51} (\bibinfo{year}{2022}).

\bibitem{Hong_Kohne_Morrell_Wang_Weeks_2017}
\bibinfo{author}{Hong, X.}, \bibinfo{author}{Kohne, M.},
  \bibinfo{author}{Morrell, M.}, \bibinfo{author}{Wang, H.} \&
  \bibinfo{author}{Weeks, E.~R.}
\newblock \bibinfo{title}{Clogging of soft particles in two-dimensional
  hoppers}.
\newblock \emph{\bibinfo{journal}{Physical Review E}}
  \textbf{\bibinfo{volume}{96}}, \bibinfo{pages}{062605}
  (\bibinfo{year}{2017}).

\bibitem{Gorkov_1962}
\bibinfo{author}{Gor'kov, L.~P.}
\newblock \bibinfo{title}{On the forces acting on a small particle in an
  acoustical field in an ideal fluid}.
\newblock \emph{\bibinfo{journal}{Soviet Physics Doklady}}
  \textbf{\bibinfo{volume}{6}}, \bibinfo{pages}{773} (\bibinfo{year}{1962}).

\bibitem{Silva_Bruus_2014}
\bibinfo{author}{Silva, G.~T.} \& \bibinfo{author}{Bruus, H.}
\newblock \bibinfo{title}{Acoustic interaction forces between small particles
  in an ideal fluid}.
\newblock \emph{\bibinfo{journal}{Physical Review E}}
  \textbf{\bibinfo{volume}{90}}, \bibinfo{pages}{063007}
  (\bibinfo{year}{2014}).

\bibitem{Lim_VanSaders_Souslov_Jaeger_2022}
\bibinfo{author}{Lim, M.~X.}, \bibinfo{author}{VanSaders, B.},
  \bibinfo{author}{Souslov, A.} \& \bibinfo{author}{Jaeger, H.~M.}
\newblock \bibinfo{title}{Mechanical properties of acoustically levitated
  granular rafts}.
\newblock \emph{\bibinfo{journal}{Physical Review X}}
  \textbf{\bibinfo{volume}{12}}, \bibinfo{pages}{021017}
  (\bibinfo{year}{2022}).

\bibitem{Wu_VanSaders_Lim_Jaeger_2023}
\bibinfo{author}{Wu, B.}, \bibinfo{author}{VanSaders, B.},
  \bibinfo{author}{Lim, M.~X.} \& \bibinfo{author}{Jaeger, H.~M.}
\newblock \bibinfo{title}{Hydrodynamic coupling melts acoustically levitated
  crystalline rafts}.
\newblock \emph{\bibinfo{journal}{Proceedings of the National Academy of
  Sciences}} \textbf{\bibinfo{volume}{120}}, \bibinfo{pages}{e2301625120}
  (\bibinfo{year}{2023}).

\bibitem{Hsiao_Lee_Samuelsen_Lipkowitz_Kronenfeld_Ilyn_Shih_Dulay_Tate_Shaqfeh_et_al._2022}
\bibinfo{author}{Hsiao, K.} \emph{et~al.}
\newblock \bibinfo{title}{Single-digit-micrometer-resolution continuous liquid
  interface production}.
\newblock \emph{\bibinfo{journal}{Science Advances}}
  \textbf{\bibinfo{volume}{8}}, \bibinfo{pages}{eabq2846}
  (\bibinfo{year}{2022}).

\bibitem{Lee_Hsiao_Lipkowitz_Samuelsen_Tate_DeSimone_2022}
\bibinfo{author}{Lee, B.~J.} \emph{et~al.}
\newblock \bibinfo{title}{Characterization of a 30 µm pixel size clip-based
  {3D} printer and its enhancement through dynamic printing optimization}.
\newblock \emph{\bibinfo{journal}{Additive Manufacturing}}
  \textbf{\bibinfo{volume}{55}}, \bibinfo{pages}{102800}
  (\bibinfo{year}{2022}).

\bibitem{Kronenfeld_Rother_Saccone_Dulay_DeSimone_2024}
\bibinfo{author}{Kronenfeld, J.~M.}, \bibinfo{author}{Rother, L.},
  \bibinfo{author}{Saccone, M.~A.}, \bibinfo{author}{Dulay, M.~T.} \&
  \bibinfo{author}{DeSimone, J.~M.}
\newblock \bibinfo{title}{Roll-to-roll, high-resolution {3D} printing of
  shape-specific particles}.
\newblock \emph{\bibinfo{journal}{Nature}} \textbf{\bibinfo{volume}{627}},
  \bibinfo{pages}{306–312} (\bibinfo{year}{2024}).

\bibitem{Lim_Jaeger_2023}
\bibinfo{author}{Lim, M.~X.} \& \bibinfo{author}{Jaeger, H.~M.}
\newblock \bibinfo{title}{Acoustically levitated lock and key grains}.
\newblock \emph{\bibinfo{journal}{Physical Review Research}}
  \textbf{\bibinfo{volume}{5}}, \bibinfo{pages}{013116} (\bibinfo{year}{2023}).

\bibitem{Krim_Yu_Behringer_2011}
\bibinfo{author}{Krim, J.}, \bibinfo{author}{Yu, P.} \&
  \bibinfo{author}{Behringer, R.~P.}
\newblock \bibinfo{title}{Stick–slip and the transition to steady sliding in
  a {2D} granular medium and a fixed particle lattice}.
\newblock \emph{\bibinfo{journal}{Pure and Applied Geophysics}}
  \textbf{\bibinfo{volume}{168}}, \bibinfo{pages}{2259–2275}
  (\bibinfo{year}{2011}).

\bibitem{Kim_Greer_2009}
\bibinfo{author}{Kim, J.-Y.} \& \bibinfo{author}{Greer, J.~R.}
\newblock \bibinfo{title}{Tensile and compressive behavior of gold and
  molybdenum single crystals at the nano-scale}.
\newblock \emph{\bibinfo{journal}{Acta Materialia}}
  \textbf{\bibinfo{volume}{57}}, \bibinfo{pages}{5245–5253}
  (\bibinfo{year}{2009}).

\bibitem{Greer_Nix_2006}
\bibinfo{author}{Greer, J.~R.} \& \bibinfo{author}{Nix, W.~D.}
\newblock \bibinfo{title}{Nanoscale gold pillars strengthened through
  dislocation starvation}.
\newblock \emph{\bibinfo{journal}{Physical Review B}}
  \textbf{\bibinfo{volume}{73}}, \bibinfo{pages}{245410}
  (\bibinfo{year}{2006}).

\bibitem{Brinckmann_Kim_Greer_2008}
\bibinfo{author}{Brinckmann, S.}, \bibinfo{author}{Kim, J.-Y.} \&
  \bibinfo{author}{Greer, J.~R.}
\newblock \bibinfo{title}{Fundamental differences in mechanical behavior
  between two types of crystals at the nanoscale}.
\newblock \emph{\bibinfo{journal}{Physical Review Letters}}
  \textbf{\bibinfo{volume}{100}}, \bibinfo{pages}{155502}
  (\bibinfo{year}{2008}).

\bibitem{Anzivino_Casiulis_Zhang_Moussa_Martiniani_Zaccone_2023}
\bibinfo{author}{Anzivino, C.} \emph{et~al.}
\newblock \bibinfo{title}{Estimating random close packing in polydisperse and
  bidisperse hard spheres via an equilibrium model of crowding}.
\newblock \emph{\bibinfo{journal}{The Journal of Chemical Physics}}
  \textbf{\bibinfo{volume}{158}}, \bibinfo{pages}{044901}
  (\bibinfo{year}{2023}).

\bibitem{Mandal_Nicolas_Pouliquen_2020}
\bibinfo{author}{Mandal, S.}, \bibinfo{author}{Nicolas, M.} \&
  \bibinfo{author}{Pouliquen, O.}
\newblock \bibinfo{title}{Insights into the rheology of cohesive granular
  media}.
\newblock \emph{\bibinfo{journal}{Proceedings of the National Academy of
  Sciences}} \textbf{\bibinfo{volume}{117}}, \bibinfo{pages}{8366–8373}
  (\bibinfo{year}{2020}).

\bibitem{Rognon_Roux_Naaïm_Chevoir_2008}
\bibinfo{author}{Rognon, P.~G.}, \bibinfo{author}{Roux, J.-N.},
  \bibinfo{author}{Naaïm, M.} \& \bibinfo{author}{Chevoir, F.}
\newblock \bibinfo{title}{Dense flows of cohesive granular materials}.
\newblock \emph{\bibinfo{journal}{Journal of Fluid Mechanics}}
  \textbf{\bibinfo{volume}{596}}, \bibinfo{pages}{21–47}
  (\bibinfo{year}{2008}).

\end{thebibliography}
%% if required, the content of .bbl file can be included here once bbl is generated
%%\input sn-article.bbl

\end{document}